\documentclass[aps]{revtex4}
\usepackage{graphicx}
\usepackage{bm}
\usepackage{amsmath}
\usepackage[T1]{fontenc}
\usepackage{mathrsfs,ae,dsfont}
\usepackage{amssymb}

\begin{document}

\title{Structure Group and Fermion-Mass-Term in General Nonlocality}

\author{Lei Han, Hai-Jun Wang}
\address{Center for Theoretical Physics and School of Physics, Jilin
University, Changchun 130012, China}

\begin{abstract}
In our previous work [J. Math. Phys. 49, 033513 (2008)] two
problems remain to be resolved. One is that we lack a minimal
group to replace GL(4,C), the other is that the Equation of Motion
(EoM) for fermion has no mass term. After careful investigation we
find these two problems are linked by conformal group, a subgroup
of GL(4,C) group. The Weyl group, a subgroup of conformal group,
can bring about the running of mass, charge etc. while making it
responsible for the transformation of interaction vertex. However,
once concerning the generation of the mass term in EoM, we have to
resort to the whole conformal group, in which the generators
$K_\mu $ play a crucial role in making vacuum vary from space-like
(or light-cone-like)to time-like. Physically the starting points
are our previous conclusion, $\vec E^2-\vec
B^2\neq 0$ for massive bosons, and the two-photon process yielding $%
e^{+}e^{-}$ pair. Finally we get to the conclusion that the mass term of
strong interaction is linearly relevant to (chromo-)magnetic flux as well as
angular momentum.
\end{abstract}

\maketitle

\section{Introduction}

In our previous paper ~\cite{WA08}, a complex-geometry approach to quantum
field theory was proposed, aiming at describing nonlocal interaction (in the
low-energy limit) through curvature effects in certain kinds of complex
space. In such approach, the field equations for interacting fermions and
bosons are obtained based on the assumption that there always exists a
complex reference frame for a fermion in which it behaves as a plane wave,
i.e. as a free (non-interacting) particle. This assumption is regarded as
''principle of nonlocality''------but actually it is a ''complexification''
of the Einstein's Equivalence Principle (EEP). The equation of motion (EoM)
for fermions is derived by demanding that the fermions move just along a
''geodesic line'' in the complex space. And by a transformation of
observers, we conclude that it is just the quadratic form of the standard
Dirac equation, yet without the mass term. The field equation for bosons is
also derived, in which the mass of boson occurs automatically and is
proportional to $\vec E^2-\vec B^2$. The conclusion holds apparently for
photon since $\vec E^2-\vec B^2=0$ there. The theory might be useful in
studying non-perturbative dynamical systems, such as hadron spectra or
scattering between hadrons. In such a theory, a few problems remain to
resolve, one of which is that the structure group GL(4,C) is too large.
Another is that a mass term is missing while deriving EoM for fermions. In
this paper we try to analyze and resolve these two problems consistently.

~~\\

In our model, there are two kinds of curving------hyperbolic curving and
elliptic curving. They are represented by two different structure groups
GL(n,C) and U(n,C)~\cite{WA08}, and in this paper we are mainly concerned
about the group GL(n,C). The 4-dimension Conformal Group is a promising
mini-group for our model, in which the scaling transformation is a typical
one in generators, nevertheless is seldom used independent of other
generators~\cite{Yu13}. The algebra structure of conformal group has been
investigated thoroughly from different aspects ~\cite
{Cartan37,Ma99,Duval0911,Henkel94,Henkel02,Henkel03,Jackiw72,Hor10}, and its
applications to quantum fields have also been widely considered. However
none of the applications is satisfactory because hitherto no other perfect
quantum system than photon field \cite{Bate1910,Cunn1910} has been found so
that the corresponding Lagrangian is conformally invariant, unless the
masses of involved particles are null~\cite{Kast08, Yu13, Jose88, Gross70,
Dirac35,Lus75}. Furthermore, according to Noether's theorem if a Lagrangian
is conformally invariant, then the trace of the energy-momentum tensor
should be null~\cite{Kast08,Yu13}. Other efforts were also experienced to
search for invariant fermion equation or scattering amplitude ~\cite
{Dirac35, Gross70, Mack69}, and even to apply it to nonlocal action ~\cite
{Ryder74,Whe4549}. In recent years there have been surges using conformal
invariance accompanied with chiral symmetry breaking in studying
hadron-level strong interaction~\cite{Ma13}. So far it can almost be
concluded that the conformal invariance is unreachable to known material
fields, by which we may suspect the applicability of conformal group.
However from another point of view, the perturbative renormalization results
~\cite{Wang11,Su99,Wang04,Cal70,Syman70} shed much light on the relationship
between the physical constants and scale transformation, that is, some of
the constants vary regularly with the energy scale. It follows that the
scale transformation can cause the variation of coupling constants and
masses. That suggests that scaling transformation might be helpful in
generating mass term in EoM mathematically. Whereas after delicate analysis,
we find out that in our model, without the help of other generators of
conformal group, the mass term cannot be generated solely by scaling
transformation.

~~\\

We analyze the mass term problem from two respects, the physical respect and
math respect. On physical side, the mass generation ascribes to dynamical
chiral symmetry breaking (DCSB) and Higgs mechanism. The effect of these two
mechanisms is to turn vacuum or a massless particle from space-like (or
light-cone like) to time-like, and thus makes particle massive. In view of
our former result that a massive boson meets the condition $\vec E^2-\vec
B^2\neq 0$, and of the Higgs mechanism that demands bosons occur first and
then decay into fermions, we employ a shortcut process $\gamma +\gamma
\longrightarrow e^{+}e^{-}$ ~\cite{Piaz09, Jiang11, Drei12, Tito12, Bakm08,
Li12, Brodsky06, Kohl14, Blinne14, Jans13, LiA14, Fedotov13} to analyze the
transition from space-like (or light-cone like) to time-like. In the process
we perceive, at least equivalently, one photon is stimulated by the other
then becomes time-like and succeedingly decays into electron pairs. The
stimulation might lead to $\vec E^2-\vec B^2\neq 0$. Such processes are
truly relevant to our consideration due to the fact that Higgs boson was
first evidenced by decaying into two-photon resonance. Such seemingly rough
physics consideration is helpful in our treating on math side, where we will
note only particular generators of conformal group are responsible for
shifting space-time or energy-momentum from space-like to time-like~\cite
{Ryder74A}, which is not able to be realized solely by the scaling
transformation of conformal group. With the above cognition we find out the
answer to mass term problem points to one conjecture in our original paper,
i.e an integral which we had not carried out.

~~\\

The structure of this paper is as follows. In sect. II we introduce two
representations of 4-dimensional conformal group. The focus is on how to
derive the spinor representation, which is actually employed as our nonlocal
structure group. In section III we link the two representations of conformal
group by physical interaction vertices in the conventional sense, especially
we elucidate the function of generators $K_\mu $ on chiral-like vertex $%
\gamma _\mu (1-\gamma _5)$, which is useful for our further discussion on
mass term in EoM. Section IV is dedicated to elaborating the mathematical
processes of massless particles (bosons and fermions) getting massive and to
deriving the mass term. Then follows the conclusions and discussions. In
appendix A. we explain why a massive fermion spinor cannot be generated
solely by the scaling transformation. The scaling transformation just
transforms a spinor from helicity representation to spinor representation.

\section{The Spatial and Spinor Representations for Conformal Group}

In this section we separately present the spatial/operator representation
and unitary/spinor representation of the 4-dimensional Conformal group~\cite
{Budi79,Yu13}, including their generators and commutations among them. The
spatial representation is mainly referencing to that of Ref.~\cite{Budi79A,
Budi79} and the unitary representation is derived by applying Cartan method ~%
\cite{Cartan37} to $SO(6)-SU(4)$ transform. The unitary representation is
the focus of this section.

We start with the null vector space (Euclidean space, with the first and the
sixth imaginary),
\begin{equation}
\sum_{a=1}^6\eta _a^2=0\text{ .}  \label{LengthinReal}
\end{equation}
reserving which gives the popular definition of conformal group~\cite
{Cartan37}. A special expression of the differential forms in 4-dimension
spatial representation can be derived directly from the above equation. In
derivation we need to apply the following variables \cite{Budi79}
\begin{equation}
x_\mu =\frac{\eta _\mu }K\text{ , where }K=\eta _5+i\,\eta _6\text{ , where }%
\mu =1,2,3,4  \label{projection}
\end{equation}
together with the differential form
\begin{equation}
\frac \partial {\partial \eta _a}=\frac 1K\{[\delta _{a\mu }-(\delta
_{a5}+i\delta _{a6})x_\mu ]\frac \partial {\partial x_\mu }+(\delta
_{a5}+i\delta _{a6})K\frac \partial {\partial K}\}\text{ , where }%
a=1,2,\cdots ,6  \label{aa}
\end{equation}
to the definition of 6-dimensional angular-momentum
\begin{equation}
M_{ab}=i(\eta _a\frac \partial {\partial \eta _b}-\eta _b\frac \partial
{\partial \eta _a})\text{, where }a,b=1,2,\cdots ,6\text{ .}  \label{bb}
\end{equation}
Then one gets the following generators for conformal group \cite{Budi79} [of
which in eq. (56)]
\begin{eqnarray}
D &=&iM_{56}=-(\eta _5\frac \partial {\partial \eta _6}-\eta _6\frac
\partial {\partial \eta _5})=i(x_\mu \frac \partial {\partial x_\mu }-K\frac
\partial {\partial K})\text{,}  \nonumber \\
P_\mu &=&M_{5\mu }+iM_{6\mu }=i\frac \partial {\partial x_\mu }\text{ , }%
K_\mu =M_{5\mu }-iM_{6\mu }=i\{-x^2\frac \partial {\partial x_\mu }+2x_\mu
x_\nu \frac \partial {\partial x_\nu }-2K\,x_\mu \frac \partial {\partial
K}\}\text{,}  \label{cc}
\end{eqnarray}
The projected form (making $K$ as constant boundary of Minkowski space\cite
{Ma99}) shifting to Minkowski convention then is
\begin{eqnarray}
D &=&i\,x_\mu \frac \partial {\partial x_\mu }\text{, }M_{\mu \nu }=i(x_\mu
\frac \partial {\partial x^\nu }-x_\nu \frac \partial {\partial x^\mu })%
\text{,}  \nonumber \\
P_\mu &=&i\frac \partial {\partial x^\mu }\text{ , }K_\mu =-i(x^2\frac
\partial {\partial x^\mu }-2x_\mu x^\nu \frac \partial {\partial x^\nu })%
\text{,}  \label{ccc}
\end{eqnarray}
where $M_{\mu \nu }$ represent the components of conventional angular
momentum in 4-dimension. The corresponding commutation relation can be
obtained by direct computation,
\begin{eqnarray}
\lbrack M_{\mu \nu },M_{\rho \sigma }] &=&i(g_{\nu \rho }M_{\mu \sigma
}+g_{\mu \sigma }M_{\nu \rho }-g_{\mu \rho }M_{\nu \sigma }-g_{\nu \sigma
}M_{\mu \rho }),  \nonumber \\
\lbrack M_{\mu \nu },P_\rho ] &=&i(g_{\nu \rho }P_\mu -g_{\mu \rho }P_\nu ),
\nonumber \\
\lbrack D,P_\mu ] &=&-iP_\mu \text{, }[D,K_\mu ]=iK_\mu ,  \nonumber \\
\lbrack D,M_{\mu \nu }] &=&0  \nonumber \\
\ \ [M_{\mu \nu },K_\rho ] &=&i(g_{\nu \rho }K_\mu -g_{\mu \rho }K_\nu )
\label{dd} \\
\ [P_\mu ,K_\rho ] &=&-2\,i\,(g_{\mu \rho }\,D+M_{\mu \rho })
\end{eqnarray}

Before using Cartan method to achieve its unitary representation, let's
review first the steps of Cartan method with $SO(3)-SU(2)$ mapping as an
example \cite{Cartan37}[of which in pp. 41-48]. To keep the invariance of $%
x_1^2+x_2^2+x_3^2=0$, one defines the matrix
\begin{equation}
X=\left(
\begin{array}{cc}
x_3 & x_1-i\,x_2 \\
x_1+i\,x_2 & -x_3
\end{array}
\right) \text{ .}  \label{cc}
\end{equation}
The trace Tr$(X^{\dagger }X)$ is $x_1^2+x_2^2+x_3^2$. With $U$ as an element
of $SU(2)$ group, we define
\begin{equation}
X^{\prime }=U^{-1}XU\text{ ,}  \label{ee}
\end{equation}
immediately we have
\begin{equation}
Tr(X^{\prime \dagger }X^{\prime })=Tr(X^{\dagger }X)\text{ ,}  \label{xx}
\end{equation}
thus $SU(2)$ group keeps the trace invariant, and by this way the group also
keeps the metric $x_1^2+x_2^2+x_3^2$. With the knowledge that the $SO(3)$
group directly reserves the metric $x_1^2+x_2^2+x_3^2$, we conclude that
Cartan matrix $X$ acts as a mapping between $SO(3)$ and $SU(2)$. By the
Cartan Matrix $X$, one can define spinor $\left(
\begin{array}{c}
\xi _0 \\
\xi _1
\end{array}
\right) $ by
\begin{equation}
X\,\left(
\begin{array}{c}
\xi _0 \\
\xi _1
\end{array}
\right) =0\text{ ,}  \label{ks}
\end{equation}
with the solution $\xi _0=\pm \sqrt{\frac{x_1-i\,x_2}2}$ and $\xi _1=\pm
\sqrt{\frac{-x_1-i\,x_2}2}$, and the reverse yields
\begin{eqnarray}
x_1 &=&\xi _0^2-\xi _1^2  \nonumber \\
x_2 &=&i\,(\xi _0^2+\xi _1^2)  \nonumber \\
x_3 &=&-2\xi _0\xi _1\text{ ,}  \label{spinor}
\end{eqnarray}
which automatically satisfies $x_1^2+x_2^2+x_3^2=0$ from which we can define
the spinor reversely.

From the above Cartan matrix $X$ we can extract the Pauli matrices $\sigma _1
$, $\sigma _2$, $\sigma _3$ separately from the coefficients of $x_1$, $x_2$%
, $x_3$. Meanwhile Pauli matrices $\sigma _1$, $\sigma _2$, $\sigma _3$ act
as the generators of $SU(2)$ group mentioned above. Furthermore it is easy
to test that $SU(2)$ group reserves the metric
\begin{equation}
\mid \xi _0\mid ^2+\mid \xi _1\mid ^2=\Xi ^{\dagger }\Xi \text{ .}
\label{Dspinor}
\end{equation}
And coincidentally the $n-$vectors form (defined in eq. (\ref{k-vector}))
based on Pauli matrices don't generate new matrices, neither the
multiplications nor the commutations among them, because they themselves are
closed. Now in what follows we would find the corresponding Cartan matrix
from $SO(6)$ to $SU(4)$/$SU(2,2)$, namely the spinor representation for
4-dimension Conformal group.

To achieve its unitary/spinor representation in 4-dimension, mimicking the
relationship between the metric $x_1^2+x_2^2+x_3^2$ and that in Eq. (\ref
{Dspinor}), we shall associate the metric in Eq. (\ref{LengthinReal}) with
the invariant quadratic form
\begin{equation}
\mid z_1\mid ^2+\mid z_2\mid ^2+\mid z_3\mid ^2+\mid z_4\mid ^2=Z^{\dagger }Z%
\text{ ,}  \label{UniLength}
\end{equation}
by the following matrix~~\cite{Este64},
\begin{equation}
A=\left(
\begin{array}{cccc}
0 & x_1+i\,x_2 & x_3+i\,x_4 & x_5+i\,x_6 \\
-(x_1+i\,x_2) & 0 & x_5-i\,x_6 & -x_3+i\,x_4 \\
-(x_3+i\,x_4) & -x_5+i\,x_6 & 0 & x_1-i\,x_2 \\
-(x_5+i\,x_6) & x_3-i\,x_4 & -x_1+i\,x_2 & 0
\end{array}
\right) \text{ .}  \label{MatrixCorres}
\end{equation}
Count the degrees of freedom of the groups conserving separately Eq. (\ref
{LengthinReal}) and Eq. (\ref{UniLength}), one finds they are both 15. Next
we only need to extract the coefficients before $x_i$'s to get the unitary
matrices as generators of $SU(4)$, just like the method used in three
dimension example~Eq (\ref{dd}-\ref{spinor}). If we want to get the
generators of $SU(2,2)$ we need only to change the signs before $x_1$ and $%
x_2$ and those ahead of corresponding matrices, which would change the eqs. (%
\ref{LengthinReal}) and (\ref{UniLength}) to
\begin{equation}
-x_1^2-x_2^2+x_3^2+x_4^2+x_5^2+x_6^2=0\ \text{ .}  \label{LengthinRealB}
\end{equation}
and
\begin{equation}
-\mid z_1\mid ^2-\mid z_2\mid ^2+\mid z_3\mid ^2+\mid z_4\mid ^2=Z^{\dagger
}Z\text{ .}  \label{UniLengthB}
\end{equation}
the latter falls into Dirac spinor like
\[
\tilde \psi =(z_1,z_2,z_3,z_4)\text{ .}
\]
It can be examined that the matrix $A$ in Eq. (\ref{MatrixCorres}) meets the
invariant expression
\begin{equation}
Tr(A^{\dagger }A)=4(x_1^2+x_2^2+x_3^2+x_4^2+x_5^2+x_6^2)  \label{trace}
\end{equation}
just like the above 3-dimension example, while the $SU(4)$ group keeps the
above trace $x_1^2+x_2^2+x_3^2+x_4^2+x_5^2+x_6^2=constant$, it
simultaneously reserves the metric Eq. (\ref{UniLength}). The above method
of linking real metric to a matrix is closely analogous to the Cartan method
of constructing a spinor representation in any real space. Actually, the
true spinor space for 4-d conformal group following Cartan method should be
of 8-dimension instead of 4-dimension \cite{Cartan37}[of which in pp.
88-89]. In what follows we would take over the process of deriving all of
the $n$-vectors along the Cartan method~\cite{Cartan37}[of which in
pp.81-83], though we work in 4-dimension rather than 8-dimension. First we
extract the matrices before $x_i$'s in Eq. (\ref{MatrixCorres}) , $i.e.1-$%
vectors,
\begin{eqnarray}
B_1 &=&\left(
\begin{array}{cc}
i\,\sigma _2 & 0 \\
0 & i\,\sigma _2
\end{array}
\right) \text{,}\;B_2=\left(
\begin{array}{cc}
-\,\sigma _2 & 0 \\
0 & \,\sigma _2
\end{array}
\right) \text{ ,}  \nonumber  \label{1vector} \\
B_3 &=&\left(
\begin{array}{cc}
0 & \,\sigma _3 \\
-\,\,\sigma _3 & 0
\end{array}
\right) \text{,}\;B_4=\left(
\begin{array}{cc}
0 & i\,I \\
-i\,I & \,0
\end{array}
\right) \text{ ,}  \nonumber \\
B_5 &=&\left(
\begin{array}{cc}
0 & \,\sigma _1 \\
-\,\,\sigma _1 & 0
\end{array}
\right) \text{,}\;B_6=\left(
\begin{array}{cc}
0 & \,-\sigma _2 \\
-\,\,\sigma _2 & 0
\end{array}
\right) \text{ .}  \label{1vector}
\end{eqnarray}
where $\sigma _i$'s are Pauli matrices. The definition of $k$-vector is
\begin{equation}
B_{k-vector}=\sum_P(-1)^PB_{n_1}B_{n_2}\cdots B_{n_k}\text{ ,}
\label{k-vector}
\end{equation}
where $P$ denotes different permutations. Apply the above formula to
2-vector, and use the corresponding subscripts to denote the $1$-vectors
involved, then
\[
B_{12}=B_1B_2-B_2B_1=\left(
\begin{array}{cc}
i\,\sigma _2 & 0 \\
0 & i\,\sigma _2
\end{array}
\right) \left(
\begin{array}{cc}
-\,\sigma _2 & 0 \\
0 & \,\sigma _2
\end{array}
\right) -\left(
\begin{array}{cc}
-\,\sigma _2 & 0 \\
0 & \,\sigma _2
\end{array}
\right) \left(
\begin{array}{cc}
i\,\sigma _2 & 0 \\
0 & i\,\sigma _2
\end{array}
\right) =0\text{ .}
\]
Similarly, let's exhaust all possibilities, then obtain other nontrivial
2-vectors
\begin{eqnarray}
B_{13} &=&2\left(
\begin{array}{cc}
0 & -\,\sigma _1 \\
\,\,\sigma _1 & 0
\end{array}
\right) \text{ , }B_{15}=2\left(
\begin{array}{cc}
0 & \,\sigma _3 \\
\,-\,\sigma _3 & 0
\end{array}
\right) \text{, }  \nonumber  \label{sigma1} \\
B_{35} &=&2i\left(
\begin{array}{cc}
\,\sigma _2 & 0 \\
0 & \,\sigma _2
\end{array}
\right) \text{, }B_{36}=2i\left(
\begin{array}{cc}
\,\sigma _1 & 0 \\
0 & \,-\sigma _1
\end{array}
\right) \text{, }  \nonumber \\
B_{46} &=&-2i\left(
\begin{array}{cc}
\,\sigma _2 & 0 \\
0 & \,-\sigma _2
\end{array}
\right) \text{, }B_{24}=2i\left(
\begin{array}{cc}
0 & \,\sigma _2 \\
\,\,\sigma _2 & 0
\end{array}
\right) \text{ ,}  \nonumber  \label{sigma1} \\
\text{ }B_{23} &=&-2i\left(
\begin{array}{cc}
0 & \,\sigma _1 \\
\,\,\sigma _1 & 0
\end{array}
\right) \text{. }  \label{sigma1}
\end{eqnarray}
We note that the new ones which are independent of $B_i$'s are just $B_{23}$
and $B_{36}$. The same line can be followed to carry out the 3-vectors.
Ignoring the repeating ones, we find the new 3-vectors independent of both
1-vectors and 2-vectors are
\begin{eqnarray}
B_{123} &\sim &\left(
\begin{array}{cc}
0 & \,\sigma _3 \\
\,\,\sigma _3 & 0
\end{array}
\right) \text{, }B_{134}\sim \left(
\begin{array}{cc}
\,\sigma _1 & 0 \\
0 & \,\sigma _1
\end{array}
\right) \text{, }  \nonumber \\
B_{145} &\sim &\left(
\begin{array}{cc}
\,\sigma _3 & 0 \\
0 & \,\sigma _3
\end{array}
\right) \text{, }B_{245}\sim \left(
\begin{array}{cc}
\,\sigma _3 & 0 \\
0 & -\,\sigma _3
\end{array}
\right) \text{, }  \nonumber \\
B_{345} &\sim &\left(
\begin{array}{cc}
0 & \,\sigma _2 \\
\,-\,\sigma _2 & 0
\end{array}
\right) \text{, }B_{146}\sim \left(
\begin{array}{cc}
I & \,0 \\
0 & -I
\end{array}
\right) \text{, }  \nonumber \\
B_{124} &\sim &\left(
\begin{array}{cc}
0 & \,I \\
\,\,I & 0
\end{array}
\right) \text{ .}  \label{sigma2}
\end{eqnarray}
Computing the 4-vectors and the higher ones would not give new independent
matrices. Finally, we can rearrange all above k-vector-produced matrices as
follows ~\cite{Este64},
\begin{eqnarray}
U_i &=&\frac 12\left(
\begin{array}{cc}
\,\sigma _i & 0 \\
0 & \,\sigma _i
\end{array}
\right)   \nonumber \\
V_\mu  &=&-\frac 12\left(
\begin{array}{cc}
\,\sigma _\mu  & 0 \\
0 & \,-\sigma _\mu
\end{array}
\right)   \nonumber \\
W_\mu  &=&\frac i2\left(
\begin{array}{cc}
\,0 & \sigma _\mu  \\
\sigma _\mu  & \,0
\end{array}
\right)   \nonumber \\
Y_\mu  &=&\frac 12\left(
\begin{array}{cc}
\,0 & \sigma _\mu  \\
-\sigma _\mu  & \,0
\end{array}
\right)   \label{sigma3}
\end{eqnarray}
where $\sigma _i$, $i=1,2,3$ are normal Pauli matrices and $\sigma _0=\left(
\begin{array}{cc}
1 & 0 \\
0 & \,1
\end{array}
\right) $. The convention can be changed from Minkowski to Euclidean spaces
while alternatively requiring $\sigma _\mu ^2=-1$, i.e. making $\sigma _0=i$
and replacing definition of $\sigma _i$ by those in \cite{Este64}. The route
of inquiring the concrete matrices following Cartan method as above could be
a shortcut that rarely mentioned in literature. It is can be checked that
the commutations among $U_i$, $V_\mu $, $W_\mu $, $Y_\mu $ are just those
for conformal group \cite{Budi79, Budi79A}, accordingly the mapping from
these matrices to differential forms turns out to be
\begin{eqnarray}
U_i &\leftrightarrow &\gamma _i\gamma _j\longrightarrow i(x_j\frac \partial
{\partial x^k}-x_k\frac \partial {\partial x^j})\longrightarrow M_{jk}
\nonumber \\
W_i &\leftrightarrow &\gamma _0\gamma _i\longrightarrow i(x_i\frac \partial
{\partial x^0}-x_0\frac \partial {\partial x^i})\longrightarrow M_{0k}
\nonumber \\
W_0 &\leftrightarrow &\gamma _5\longrightarrow i\,x_\mu \frac \partial
{\partial x_\mu }\longrightarrow D  \nonumber \\
V_\mu +Y_\mu  &\leftrightarrow &\gamma _\mu (1-\gamma _5)\longrightarrow
i\frac \partial {\partial x^\mu }\longrightarrow P_\mu   \nonumber \\
V_\mu -Y_\mu  &\leftrightarrow &\gamma _\mu (1+\gamma _5)\longrightarrow
-i(\frac 12x_\nu x^\nu \frac \partial {\partial x_\mu }-x_\mu x_\nu \frac
\partial {\partial x_\nu })\text{ }\longrightarrow K_\mu \text{.}
\label{mapping}
\end{eqnarray}
We use $\longrightarrow $ to represent the accurate mappings and $%
\leftrightarrow $ the equivalence. After examining the commutations by
computer, we have to extend the matrix of operator $D$ to be $1+\gamma _5$,
due to the one between $P_\mu $ and $K_\mu $.

\section{The physical relationship between the two representations of
conformal group}

~~\\

In this section we use scaling transformation as paradigm to investigate the
affection of conformal group on interaction vertex. Enlightened by applying
Lorentz transformation to Dirac equation, we first try to link physically
the spatial form of scaling transformation with its spinor/unitary form, the
former representing the realistic expansions and contractions (dilatation
and shrinkage) of space-time, the latter representing the intrinsic freedom
closely analogous to spin angular momentum. Then we will generalize the
results from scaling transformation to other generators of conformal group.
By this way we analyze the structure of conformal group and its relationship
with the vertices of conventional quantum field theory, and judge the
concrete physical significance of every block of the conformal group.

~~\\

As for Lorentz transformation, the transformation matrix $\left( \Lambda
_{\;\,\mu }^\nu \right) $ for $j^\mu (y)=\bar \psi (y)\gamma ^\mu \psi (y)$
corresponds to a complex transformation $S$ for $\psi (y)$ so that the
effect of the transformed result $\bar \psi (y)S^{-1}\gamma ^\mu S\psi (y)$
is equivalent to $\bar \psi (y)\Lambda _{\;\,\nu }^\mu \gamma ^\nu \psi (y)$%
. Referencing the case of Lorentz transformation, our goal in this section
is to search for the corresponding vertex-form $\Gamma ^\mu $ so that it
links with transformation $S^{\prime }$ by $S^{\prime -1}\Gamma ^\mu
S^{\prime }=\Lambda _{\;\,\nu }^{\prime \mu }\Gamma ^\nu $, where $S^{\prime
}=e^{\frac u2(1+\gamma _5)}$, $1+\gamma _5$ is the spinor representation of
the scaling operator $D$, and $\Lambda _{\;\,\nu }^{\prime \mu }$ represent
tensor's components of scaling transformation. The similar method was used
in a previous paper \cite{WA08B}, but for totally different motivation.

~~\\

Usually we perform the spatial Lorentz transformation on the vectors $A_\mu $
and $\gamma ^\mu $. Obviously this combination brings about invariant
formalism like $A^\nu (q^2)\bar \psi (p)\gamma _\nu \psi (p^{\prime })$. We
follow the convention that the same set of $\{\gamma ^\mu \}$ is used in
different coordinate systems, which naturally yields an equivalence
transformation $S$ satisfying~\cite{Mandl10,Feynman62}
\begin{equation}
S^{-1}\gamma ^\mu S=\Lambda _{\;\nu }^\mu \gamma ^\nu \,=\gamma ^{\prime \mu
},  \label{EquTran1}
\end{equation}
where $\Lambda _{\;\nu }^\mu $ stand for the tensors' components of the
Lorentz transformation. Substituting Eq.(\ref{EquTran1}) into $A_\mu (x)\bar
\psi (x)\gamma ^\mu \psi (x)$ yields
\begin{equation}
A_\mu ^{\prime }(y)\bar \psi ^{\prime }(y)S^{-1}\gamma ^\mu S\psi ^{\prime
}(y)=A_\mu ^{\prime }(y)\bar \psi (y)\gamma ^{\prime \mu }\psi (y)\text{%
\thinspace .}  \label{EquTran2}
\end{equation}

While looking for $\Gamma ^\mu $ we would follow the same convention as that
in the above paragraph, i.e., in different coordinate system we use the same
set of $\{\Gamma ^\mu \}$. Then analogously, we use the form of the above
formula Eq. (\ref{EquTran1}) for scaling transformation as
\begin{equation}
S^{\prime -1}\Gamma ^\mu S^{\prime }=\Lambda _{\;\,\nu }^{\prime \mu }\Gamma
^\nu \text{\thinspace ,}  \label{SpinorCoodA}
\end{equation}
where formally we have used $\Lambda _{\;\,\nu }^{\prime \mu }$ to represent
the scaling transformation to every coordinate component ~\cite{Dirac35,
Gross70, Mack69}[of which eq.(2)] instead of using the usual form $%
e^{-\alpha }$\cite{Budi79A}. Different from the operator $\mu \frac{\text{d}%
}{\text{d}\mu }$ appearing in renormalization group equation~\cite
{Su99,Cal70,Syman70}
\begin{equation}
\mu \frac{\text{d}\Lambda _R}{\text{d}\mu }+\gamma _F\,\Lambda _R=0\text{ ,}
\end{equation}
where $\gamma _F$ is the anomalous scaling dimension defined by
\[
\gamma _F=\mu \frac{\text{d}}{\text{d}\mu }\ln Z_F\text{ ,}
\]
here the operator $D$ has the usual form $D=i\,x^\nu \partial _\nu $ , being
a hermit one. With the relation $e^{-i\,\alpha \,D}p_\mu e^{i\,\alpha
\,D}=e^{-\alpha }p_\mu $, i.e. $[D,\,p_\mu ]=-i\,p_\mu $\cite{Budi79A}, we
have
\begin{equation}
(\Gamma ^\mu p_\mu )_{scaling\text{ }transform}^{\prime }=S^{\prime
-1}\Gamma ^\mu S^{\prime }\Lambda _{\;\,\mu }^{\prime \nu }\,p_\nu \text{%
\thinspace }=S^{\prime -1}\Gamma ^\mu S^{\prime }\;e^{-i\,\alpha \,D}p_\mu
e^{i\,\alpha \,D}\text{.}  \label{SpinorCoodB}
\end{equation}
Now let's submit $S^{\prime }=e^{\frac u2\gamma _5}$ obtained from the last
section (henceforth we use $e^{\frac u2\gamma _5}$ instead of $e^{\frac
u2(1+\gamma _5)}$ as scaling transformation while no confusion occurs),
where $u$ is the infinitesimal parameter. Formally we get
\begin{eqnarray}
S^{\prime -1}\Gamma ^\mu S^{\prime }\Lambda _{\;\,\mu }^{\prime \nu }\,p_\nu
&=&\,e^{-\frac u2\gamma _5}\Gamma ^\mu e^{\frac u2\gamma _5}(p_\mu )_{scaling%
\text{ }transform}^{\prime }  \nonumber  \label{SpinorCoodC} \\
&=&\,e^{-\frac u2\gamma _5}\Gamma ^\mu e^{\frac u2\gamma _5}e^{-i\,\alpha
\,D}p_\mu e^{i\,\alpha \,D}  \nonumber \\
&=&e^{-\frac u2\gamma _5}\Gamma ^\mu e^{\frac u2\gamma _5}e^{-\alpha }p_\mu
\text{ }  \nonumber  \label{SpinorCoodC} \\
&\doteq &\,e^{-\frac u2\gamma _5}\Gamma ^\mu e^{\frac u2\gamma _5}p_\mu
(1-\alpha )\text{ .}  \label{SpinorCoodC}
\end{eqnarray}
From the experience of calculating $\gamma -$matrix and the following
relations
\begin{equation}
e^{-\frac u2\gamma _5}\gamma ^\mu e^{\frac u2\gamma _5}\simeq (1-\frac
u2\gamma _5)\gamma ^\mu (1+\frac u2\gamma _5)\simeq \gamma ^\mu +u\,\gamma
^\mu \gamma _5\,,  \label{ConcreteA}
\end{equation}

\begin{equation}  \label{ConcreteB}
e^{-\frac u2\gamma _5}\gamma ^\mu \gamma _5e^{\frac u2\gamma _5}\simeq
(1-\frac u2\gamma _5)\gamma ^\mu \gamma _5(1+\frac u2\gamma _5)\simeq \gamma
^\mu \gamma _5+u\,\gamma ^\mu \,,
\end{equation}

\begin{equation}
e^{-\frac u2\gamma _5}\gamma ^\mu (1\pm \gamma _5)e^{\frac u2\gamma
_5}\simeq (1-\frac u2\gamma _5)\gamma ^\mu (1\pm \gamma _5)(1+\frac u2\gamma
_5)\simeq (1\pm u\,)\gamma ^\mu (1\pm \gamma _5)\,,  \label{ConcreteC}
\end{equation}
we find out a possible form of $\Gamma ^\mu $%
\begin{equation}
\Gamma ^\mu =\gamma ^\mu (1\pm \gamma _5)\text{ while }\alpha \sim u\text{ .}
\label{Important}
\end{equation}
The transformation in Eqs. (\ref{ConcreteA}, \ref{ConcreteB}, \ref{ConcreteC}%
) can be replaced by $e^{-\frac u2(1+\gamma _5)\,}$ and $e^{\frac
u2(1+\gamma _5)}$, the results would be the same obviously. The coefficients
$(1\pm u\,)$ of Eq. (\ref{ConcreteC}) can be contracted now to be $1$ with
coefficients $(1\mp u\,)$ that come from the transformation of $p_\mu $. And
we note that the infinitesimal parameters $u$ and $\alpha $ are not
independent. By this way we set up the relationship between the operator $D$
and $S^{\prime }=e^{\frac u2(1+\gamma _5)}$ directly.

And we know $S^{\prime }=e^{\frac u2(1+\gamma _5)}$ is acting on Dirac
spinor as expected. The same transformation holds for vertex $\Gamma ^\mu
\,A_\mu $, as well as $\Gamma ^\mu \,p_\mu $. The resultant vertex is
different from that of Ref. \cite{Mack69} due to the choice of $\gamma _5$,
since we have followed the convention of Quantum Field Theory. In fact we
have extended the transformation, the interaction vertex and spinor space
simultaneously, and these elements can be extended further while we
involving more generators of conformal group.

What if we perform the scaling transformation $S^{\prime }$ succeedingly $N$
times upon the vector vertex-form $\gamma ^\mu $. Different from Eqs. (\ref
{ConcreteA}, \ref{ConcreteB}, \ref{ConcreteC}), now we employ the following
formulism without approximation
\begin{equation}
(e^{-\frac u2(1+\gamma _5)})^N\gamma ^\mu (e^{\frac u2(1+\gamma
_5)})^N=\gamma ^\mu \cosh Nu+\gamma ^\mu \gamma _5\sinh Nu\,\text{ ,}
\label{Correct}
\end{equation}
from which one notes that the vector vertex arrives at its limits $\gamma
^\mu (1\pm \gamma _5)$ only if $\frac{\cosh Nu}{\sinh Nu}\rightarrow \pm 1$,
i.e. $Nu$ $\rightarrow \pm \infty $. $Nu$ $\rightarrow \pm \infty $ means
one carrying out enough steps of inflating or shrinking transformation. We
call such states involving interaction vertices $\gamma ^\mu (1\pm \gamma
_5) $ as extreme states, which evolve from the interaction vertex $\gamma
^\mu $ after the scale constantly changing. And the variation of coupling
constant is assumed to be absorbed into coupling constant. It turns out that
such scaling transformation doesn't conserve the vector-dominant
interaction, or alternatively, the transformation tends to transform the
relating spinor from a normal one to a chiral one.

~~\\

We have derived the above limits with the relation $e^{-i\,\alpha \,D}p_\mu
e^{i\,\alpha \,D}=e^{-\alpha }p_\mu $, i.e. $[D,\,p_\mu ]=-i\,p_\mu $\cite
{Budi79A}, The $P_\mu $ plays the role of the production operator or
annihilation operator of $D$ from the point of view of quantum mechanics.
Analogously, we may notice another similar relationship among conformal
group by the commutation $[M_{\mu \nu },K_\rho ]=i(g_{\nu \rho }K_\mu
-g_{\mu \rho }K_\nu )$, however we fail to get the variation of vertex $%
\gamma ^\rho (1\pm \gamma _5)$ with the transformation $e^{\frac u2\gamma
_\mu \gamma _\nu }$ (operation of $M_{\mu \nu }$ in spinor space). So the
analog cannot be followed by any other commutations of conformal group. The
details are as follows. While the angular momentum transformation is written
to be $S=e^{\frac u2\gamma _\mu \gamma _\nu }$, its effect on vector vertex $%
\gamma ^\mu $ [we label it by $\Gamma ^\mu $]
\begin{eqnarray}
\Gamma ^{\mu \prime } &=&e^{-\frac u2\gamma _\rho \gamma _\sigma }\Gamma
^\mu e^{\frac u2\gamma _\rho \gamma _\sigma }\approx \left( 1-\frac u2\gamma
_\rho \gamma _\sigma \right) \Gamma ^\mu \left( 1+\frac u2\gamma _\rho
\gamma _\sigma \right) \approx \gamma ^\mu +u\left( g_\rho ^\mu \gamma
_\sigma -g_\sigma ^\mu \gamma _\rho \right)  \nonumber \\
P_\mu ^{\prime } &=&e^{-iaM_{\rho \sigma }}P_\mu e^{iaM_{\rho \sigma
}}=P_\mu +ia\left[ P_\mu ,M_{\rho \sigma }\right] =P_\mu -a\left( g_{\mu
\rho }P_\sigma -g_{\mu \sigma }P_\rho \right) \text{ .}
\end{eqnarray}
Combining the above transformations yields
\begin{eqnarray}
\left( \Gamma ^\mu P_\mu \right) ^{\prime } &=&\left[ \gamma ^\mu +u\left(
g_\rho ^\mu \gamma _\sigma -g_\sigma ^\mu \gamma _\rho \right) \right]
\left[ P_\mu -a\left( g_{\mu \rho }P_\sigma -g_{\mu \sigma }P_\rho \right)
\right]  \nonumber \\
\ &=&\Gamma ^\mu P_\mu +u\left( g_\rho ^\mu \gamma _\sigma -g_\sigma ^\mu
\gamma _\rho \right) P_\mu -a\gamma ^\mu \left( g_{\mu \rho }P_\sigma
-g_{\mu \sigma }P_\rho \right)  \nonumber \\
\ &=&\Gamma ^\mu P_\mu +\left( u+a\right) \left( \gamma _\sigma P_\rho
-\gamma _\rho P_\sigma \right) \text{ ,}
\end{eqnarray}
when $u\sim -a$, one obtains $\left( \Gamma ^\mu P_\mu \right) ^{\prime
}=\Gamma ^\mu P_\mu $.$\ $If replacing $\gamma ^\mu $ axil-vector vertex
yeilds $\gamma ^\mu (1\pm \gamma _5)$, we would get the same results
\begin{eqnarray}
\Gamma ^{\mu \prime } &=&e^{-\frac u2\gamma _\rho \gamma _\sigma }\Gamma
^\mu e^{\frac u2\gamma _\rho \gamma _\sigma }\approx \left( 1-\frac u2\gamma
_\rho \gamma _\sigma \right) \Gamma ^\mu \left( 1+\frac u2\gamma _\rho
\gamma _\sigma \right) \approx \Gamma ^\mu +u\left( g_\rho ^\mu \Gamma
_\sigma -g_\sigma ^\mu \Gamma _\rho \right)  \nonumber \\
P_\mu ^{\prime } &=&e^{-iaM_{\rho \sigma }}P_\mu e^{iaM_{\rho \sigma
}}=P_\mu +ia\left[ P_\mu ,M_{\rho \sigma }\right] =P_\mu -a\left( g_{\mu
\rho }P_\sigma -g_{\mu \sigma }P_\rho \right) \text{ .}
\end{eqnarray}
Combining the above transformations yields
\begin{eqnarray}
\left( \Gamma ^\mu P_\mu \right) ^{\prime }=\Gamma ^\mu P_\mu +\left(
u+a\right) \left( \Gamma _\sigma P_\rho -\Gamma _\rho P_\sigma \right) \text{
.}
\end{eqnarray}

~~\\

As for the kinetic term of an extended particle in the extreme state where
the scale transformation is repeated infinite times, the momentum becomes
light-cone like and the kinetic mass tends to zero since,
\begin{equation}
m_{kinetic}^2=(\Gamma _\mu p^\mu )(\Gamma _\nu p^\nu )=\gamma _\mu (1-\gamma
_5)p^\mu \gamma _\nu (1-\gamma _5)p^\nu =p^2(1+\gamma _5)(1-\gamma _5)=0%
\text{ ,}  \label{Kmass}
\end{equation}
here the $m^2=0$ may just have comparable meaning while its momentum is very
large and its mass can be ignored approximately, we call such mass as
kinematic mass. So we conclude the extreme vertices $\gamma ^\mu (1\pm
\gamma _5)$ give rise to no mass, and to make a fermion massive, we have to
use the group method to make the interaction vertex leave the form $\gamma
^\mu (1\pm \gamma _5)$. However since that the limits $\gamma ^\mu (1\pm
\gamma _5)$ of interaction form are invariant under transformation $e^{\frac
u2(1+\gamma _5)}$, it is impossible for $e^{\frac u2(1+\gamma _5)}$ to make
the extreme states $\gamma ^\mu (1\pm \gamma _5)$ leave its form. To make a
fermion leave this state to get mass, it is necessary to consider further
the other generators of conformal group. Whereas according to our knowledge
of commutations among the generators of conformal group, there is no much
room to choose another generator. For the vertex $\gamma ^\mu (1-\gamma _5)$
we may choose the generators $K_\mu $, otherwise for $\gamma ^\mu (1+\gamma
_5)$ we may choose $P_\mu $. For example, for $\Gamma ^\mu =\gamma ^\mu
(1-\gamma _5)$ we have,
\begin{eqnarray}
\Gamma ^{\mu \prime } &=&e^{\frac u2\gamma _\rho \left( 1+\gamma _5\right)
}\gamma ^\mu \left( 1-\gamma _5\right) e^{\frac u2\gamma _\rho \left(
1+\gamma _5\right) }  \nonumber \\
\  &=&[1+\frac u2\gamma _\rho \left( 1+\gamma _5\right) ]\gamma ^\mu \left(
1-\gamma _5\right) [1-\frac u2\gamma _\rho \left( 1+\gamma _5\right) ]
\nonumber \\
\  &=&\gamma ^\mu \left( 1-\gamma _5\right) +\frac u2\left[ \gamma _\rho
\left( 1+\gamma _5\right) ,\gamma ^\mu \left( 1-\gamma _5\right) \right] -%
\frac{u^2}4\gamma _\rho \left( 1+\gamma _5\right) \gamma ^\mu \left(
1-\gamma _5\right) \gamma _\rho \left( 1+\gamma _5\right)   \nonumber \\
\  &=&\gamma ^\mu \left( 1-\gamma _5\right) +2u\left[ g_\rho ^\mu \left(
1-\gamma _5\right) -\gamma ^\mu \gamma _\rho \right] -u^2\gamma _\rho \gamma
^\mu \gamma _\rho \left( 1+\gamma _5\right) \text{ ,}
\end{eqnarray}
when $u\longrightarrow 0$, $\Gamma ^{\mu \prime }\longrightarrow \left(
\gamma ^\mu +2ug_\rho ^\mu \right) \left( 1-\gamma _5\right) -2u\gamma ^\mu
\gamma _\rho $ and when $u\longrightarrow \infty $, $\Gamma ^{\mu \prime
}\longrightarrow -u^2\gamma _\rho \gamma ^\mu \gamma _\rho \left( 1+\gamma
_5\right) $. Then combine the above equation with the spatial transformation
$P_\mu ^{\prime }=e^{-iaK_\rho }P_\mu e^{iaK_\rho }\approx \left[ 1-iaK_\rho
\right] P_\mu \left[ 1+iaK_\rho \right] =P_\mu +ia\left[ P_\mu ,K_\rho
\right] =P_\mu +2a\left( g_{\mu \rho }D-M_{\mu \rho }\right) $, we have
\begin{eqnarray}
\left( \Gamma ^\mu P_\mu \right) ^{\prime } &\approx &\left[ \left( \gamma
^\mu +2ug_\rho ^\mu \right) \left( 1-\gamma _5\right) -2u\gamma ^\mu \gamma
_\rho \right] \left[ P_\mu +2a\left( g_{\mu \rho }D-M_{\mu \rho }\right)
\right]   \nonumber \\
\  &\approx &\Gamma ^\mu P_\mu +2u\left[ g_\rho ^\mu \left( 1-\gamma
_5\right) -\gamma ^\mu \gamma _\rho \right] P_\mu +2a\gamma ^\mu \left(
1-\gamma _5\right) \left( g_{\mu \rho }D-M_{\mu \rho }\right) \text{ .}
\label{ToVec}
\end{eqnarray}
The generator $P_\mu $ would have similar effect on vertex $\gamma ^\mu
(1+\gamma _5)$. The generator $K_\mu (P_\mu )$ first makes the bosons
getting mass by transmitting space-time from space-like to time-like, then
it makes fermions (with chiral interaction vertex, meaning the fermions have
no dynamical mass) leave the vertex, i.e. makes fermions massive.

\section{Generating fermion mass term with the help of conformal group}

~~\\

Conventionally only linearly flat unitary-complex-space is used in quantum
theory. In our model we have instead tried to express some of the effects,
such as interaction or symmetries breaking etc., by curved complex-space in
the strong interaction or strong correlation regime. Once a space is curved,
particles become living in a larger space with larger symmetry group. Here
we find that the space-curving can also be helpful in understanding the mass
generation. Mathematically, we explain the mass generation in two steps. The
first is how a boson becomes massive with the help of conformal
transformation, the second is how a massless fermion gets massive from the
massive boson. Coincidentally, we find the two steps could be ascribed to
the same generators of conformal group. Furthermore, we will discuss the
first step by photons pair production \cite{Piaz09, Jiang11, Drei12, Tito12,
Bakm08, Li12, Brodsky06, Kohl14, Blinne14, Jans13, LiA14, Fedotov13}, $%
\gamma +\gamma \longrightarrow e^{+}e^{-}$. The previous works on photons
pair production mainly focus on calculating the amplitude in semi-classical
level. In this paper we don't get involved in the concrete calculation of
the amplitude of the process, but instead focus on how the process could be
in accordance with the inequality $\vec E^2-\vec B^2\neq 0$.

~~\\

To interpret the first step, let's carry some simple calculation of
energy-momentum of classical electromagnetic field. It is well known that
the energy density of electromagnetic field is
\[
\varpi =\frac 12(\vec E^2+\vec B^2)\text{ ,}
\]
and the momentum of electromagnetic fields is Poynting vector
\[
\vec S=\vec E\times \vec B\text{ ,}
\]
using the formula
\[
(\vec a\times \vec b)\times \vec c=(\vec a\cdot \vec c)\vec b-(\vec b\cdot
\vec c)\vec a\text{ ,}
\]
we obtain
\begin{eqnarray*}
\varpi ^2-\vec S^2 &=&\frac 14(\vec E^2+\vec B^2)^2-(\vec E\times \vec B)^2
\\
\ &=&\frac 14(\vec E^2+\vec B^2)^2-(\vec E\times \vec B)\cdot (\vec E\times
\vec B) \\
\ &=&\frac 14(\vec E^2+\vec B^2)^2-(\vec E\times \vec B\times \vec E)\cdot
\vec B \\
\ &=&\frac 14(\vec E^2+\vec B^2)^2-[(\vec E\cdot \vec E)\vec B-(\vec E\cdot
\vec B)\vec E]\cdot \vec B \\
\ &=&\frac 14(\vec E^4+2\vec E^2\vec B^2+\vec B^4)-(\vec E\cdot \vec E)(\vec
B\cdot \vec B) \\
\ &=&\frac 14(\vec E^4-2\vec E^2\vec B^2+\vec B^4) \\
\ &=&\frac 14(\vec E^2-\vec B^2)^2\text{ ,}
\end{eqnarray*}
which to some extent mimics the general relationship of energy and momentum $%
p_0^2-\vec p^2=m^2$. And the result $\vec E^2-\vec B^2\neq 0$ corresponds to
boson mass according to our conclusion of previous paper. We note that the
free electromagnetic field satisfies $\varpi ^2-\vec S^2=\frac 14(\vec
E^2-\vec B^2)^2=0$ since $\mid \vec E\mid =\mid \vec B\mid $ in nature unit.
And if the photons participate in the reaction $\gamma +\gamma
\longrightarrow e^{+}e^{-}$, then maybe a photon meets $\vec E^2-\vec
B^2\neq 0$, subsequently $(\vec E^2-\vec B^2)^2>0$ and thus $\varpi ^2-\vec
S^2>0$ , which is the typical time-like relationship of energy-momentum.
This suggests that somehow a photon has gained its mass by the stimulation
of another photon. In view of the above analysis, we regard one
ultra-high-energy (laser) photon stimulates another light-cone photon to
make it transit to time-like\cite{Kohl14, Blinne14, Jans13, LiA14, Fedotov13}%
. We find the conformal group can be responsible for such transition from
space-(light-cone-)like to time-like .

~~\\

The second step is to transform a massless state ''like neutrino'' to a
massive one ''like electron''~\cite{explaA}. Since even for a massive boson
it can decay into many massless energetic fermions, at least from
mathematics it is possible. With the analysis of the last section, we note
that the massless state seems to occur effectively at the limits of very
high energy while the mass of fermion is very small comparing with the
kinetic energy. At the limits, the fermions interact with each other by the
vertex $\gamma ^\mu (1\pm \gamma _5)$, coincidentally the neutrinos pick the
$\gamma ^\mu (1-\gamma _5)$. So the interaction vertex deviating from the
form $\gamma ^\mu (1-\gamma _5)$ is a sign of ''neutrino'' getting its mass.
With the analysis of the previous section, we know that the dilatation
transformation cannot play the role in varying the interaction vertex, since
$e^{\frac u2(1+\gamma _5)}$ reserves the vertex $\gamma ^\mu (1-\gamma
_5)A_\mu $ [$e^{\frac u2(1+\gamma _5)}$ can only ex-change the spinor
between helicity representation and spin representation. See Appendix A.].
Neither can the momentum transformation $M_{\mu \nu }=e^{\frac u2\gamma _\mu
\gamma _\nu }$ change the vertex $\gamma ^\mu (1-\gamma _5)$, as evidenced
in the last section. So let's turn to the transformation $e^{\frac u2K_\nu }$
, while the $e^{\frac u2K_\nu }$ keeps directly the vertex $\gamma ^\mu
(1-\gamma _5)$, it does not keep the whole form $\gamma ^\mu (1-\gamma
_5)A_\mu $ invariant, nor does the generator $P_\mu $ keep the whole form $%
\gamma ^\mu (1+\gamma _5)A_\mu $. Thus we choose $K_\mu $ as a candidate to
make $\gamma ^\mu (1-\gamma _5)A_\mu $ leave $\gamma ^\mu (1-\gamma _5)$ and
begin to run. It runs to normal interaction vertex $\gamma ^\mu $, as shown
in Eq. (\ref{ToVec}) , which is for massive fermions.

~~\\

Now let's come back to the equation of our model. In our previous paper ~%
\cite{WA08}, we have laid two ways of generating fermion mass term in EoM,
which are just the possibilities from the viewpoint of mathematics. Now with
the deep insight of physics, we recognize that the second one must be the
candidate for the mass term, i.e. calculating a remaining integral in the
equation. Now we will treat the candidate term $-i\int A_\nu \partial
_\lambda \,\partial ^\nu \psi \,dx^\lambda $ in the EoM. Considering the
hermit of the operators $i\partial _\lambda $, $\,i\partial ^\nu $, and
apart from the coefficients, we have
\begin{eqnarray}
&&\ \ \ \ \int A_\nu \partial _\lambda \,\partial ^\nu \psi \,dx^\lambda
\nonumber \\
\  &=&\int dx^\lambda \partial _\lambda \,(\partial ^\nu A_\nu )\psi \,
\nonumber \\
\  &=&\int (dt\frac \partial {\partial t}-d\vec x\cdot \vec \nabla )\,(\frac
\partial {\partial t}A_0-\vec \nabla \cdot \vec A)\psi \,\text{ .}
\end{eqnarray}
If we omit the term $\vec \nabla \cdot \vec A$ by considering the Columb
gauge as used previously \cite{WA08}, then the above equation yeilds
\begin{eqnarray}
&&\ \ \ \ \int (dt\frac \partial {\partial t}-d\vec x\cdot \vec \nabla
)\,(\frac \partial {\partial t}A_0)\psi   \nonumber \\
\  &=&\int [dt\frac{\partial ^2}{\partial t^2}A_0-\frac \partial {\partial
t}(d\vec x\cdot \vec \nabla A_0)]\psi   \nonumber \\
\  &=&\int [dt\frac{\partial ^2}{\partial t^2}A_0-d\vec x\cdot \frac
\partial {\partial t}(\vec E+\frac \partial {\partial t}\vec A)]\psi
\nonumber \\
\  &=&\int [dt\frac{\partial ^2}{\partial t^2}A_0-d\vec x\cdot \frac{%
\partial ^2}{\partial t^2}\vec A-d\vec x\cdot (\vec \nabla \times \vec
B-\vec J)]\psi   \nonumber \\
\  &=&\int [dt\frac{\partial ^2}{\partial t^2}A_0-d\vec x\cdot \frac{%
\partial ^2}{\partial t^2}\vec A-d\vec x\cdot (\vec \nabla \times \vec
\nabla \times \vec A-\vec J)]\psi   \nonumber \\
\  &=&\int [dt\frac{\partial ^2}{\partial t^2}A_0-d\vec x\cdot \frac{%
\partial ^2}{\partial t^2}\vec A-d\vec x\cdot (-\vec \nabla ^2\vec A-\vec
J)]\psi   \nonumber \\
\  &=&\int [dt\frac{\partial ^2}{\partial t^2}A_0-d\vec x\cdot (\frac{%
\partial ^2}{\partial t^2}\vec A-\vec \nabla ^2\vec A)+d\vec x\cdot \vec
J]\psi   \nonumber \\
\  &=&\int [dt\frac{\partial ^2}{\partial t^2}A_0-d\vec x\cdot \Box \vec
A+d\vec x\cdot \vec J]\psi   \nonumber \\
\  &=&\int [dt\frac{\partial ^2}{\partial t^2}A_0+m^2d\vec x\cdot \vec
A+d\vec x\cdot \vec J]\psi   \nonumber \\
\  &=&(\lambda _1m^2\phi +\lambda _2L)\psi \text{ ,}
\end{eqnarray}
where we propose $\psi $ is almost flat (thus unchanged) under perturbative
condition, $\phi $ is a flux relating to concrete situation, and $L$ is an
angular moment. The imaginary number $i$ is absorbed by $d\vec x$ since we
have carried the conformal transformation operated by $K_\mu $, which makes
space-time varying from space-like to time-like, i.e. isotropically
interpreted by $x_\mu \rightarrow ix_\mu $~\cite{Ryder74}. In the last two
equalities of eq. (45) we have used the equation referring to (9.27) of
previous paper \cite{WA08}
\begin{equation}
\vec A\cdot \Box \vec A=\vec B^2-\vec E\stackrel{d}{=}-m^2\vec A^2\text{ ,}
\end{equation}
where the equation holds only while the considered transverse filed (Coulomb
gauge) has no source, thus no scalar component \cite{Mandl10} (of first
chapter). Now we discuss the mass term problem under same situ, so follows
the same conditions. The conditions seem not so general, however at least
they reflect certain aspects of physics, actually the Coulomb gauge $\nabla
\cdot \vec A=0$ plus radiation fields method can resolve a wide class of
pratical problems.

To understand the $\lambda _1$ and $\lambda _2$ physically, let's analyze
the first term and the second term in the last equality of eq. (45). The
first term is more like for hadrons, or for another kind of composite
particle. At first sight, the square of fermion mass being proportional to
the square of boson mass looks strange. However, this might become possible
if we look into Higgs-mechanism \cite{Mandl10} (eqs. (19.8) and (19.4)),
with relations $m_l=\upsilon g_l/\sqrt{2}$ , and $m_W=\frac 12\upsilon g$,
where $g$ and $g_l$ being adjusting parameters and $\upsilon $ a part of a
scalar fields. The two relations suggest the reasonability of the formula
like $m_{fermion}^2\propto m_{boson}^2$. On the other hand, we can analyze
the dimension of the first term. Concerning the geometrical phase factor
like $\exp [i\frac e\hbar \oint \vec A\cdot d\vec r]$, the dimension of
magnetic flux is $\left[ \frac \hbar e\right] $. And concerning the
definition of magnetic moment $\vec \mu =\frac e{m_f}\vec S$ ($\vec S$ is
the spin, $m_f$ is the fermion mass), the dimension of magnetic moment is $%
\left[ \frac{\hbar e}{m_f}\right] $. Therefore the ratio of the flux and
magnetic moment has dimension like $\left[ \frac \hbar e/(\frac{\hbar e}{m_f}%
)\right] =\left[ \frac{m_f}{e^2}\right] $, i. e. if alternatively we write
the first term as $\lambda _1^{\prime }\frac \phi \mu $ (where $\mu $ is the
absolute value of magnetic moment) then the $\left[ \lambda _1^{\prime
}=\lambda _1m\hbar e\right] $ is the proportional coeficient of flux and
magnetic moment, which is coincident with the view that this mass term is
for a composite particle. We also observe that in natural unit $\lambda
_1^{\prime }$ differ from $\lambda _1$ in a mass factor, where $\lambda _1$
itself has the dimension reciprocal of magnetic flux. Please caution that
although we constantly mention the electric or magnetic fields, they are not
the true electromagnetic fields. We just employ the analorgy between Abelian
field and non-Abelian field, our interest is non-Abelian field.

The second term may be for an elementary particle. The term mimics
the result $M^2\propto \mu ^2L+\alpha $ ($\mu ^2$, $\alpha $
constants) from Regge poles~\cite{Mas12}. From this analogy the
parameter $\lambda _2$ has the dimension of mass square, since
angular momentum $L$ is dimensionless. At present stage we may
further surmise that the first term, which represents the mass of
a hadron fermion and thus is dependent on interaction, may be
related to DCSB~\cite{Chang11,Kor14}. And the second term may
possibly link with our suggested process $\gamma +\gamma
\longrightarrow e^{+}e^{-}$, in which the photon in curved complex
space is split into elementary particles. At the same time the
electro-field and magnetic field are wrapped: the electro-field
wrapping to form charges, and the magnetic field wrapping to form
spins relating to masses. Thus both the first and second terms of
the equation correspond to similar pictures, the picture of curved
(wrapped) complex-space. This conclusion may help us make the mass
problem calculable, not just as legitimate parameters confirmed by
DCSB and Higgs mechanism. And also we should caution that here the
obtained mass may be just ''effective mass'', since classically
there is no system mimicking the strong interaction and strong
correlation, and all that we have at hand are just the experiences
of treating strong interaction in previous literature
~\cite{Hooft81, Bao07, Bao13}, which must be relevant to topology.
And mostly topology is relating to magnetic flux.

\section{Conclusions and Discussions}

In this paper we have analyzed the effect of conformal group on interaction
vertices, especially on the vector vertex and the chiral vertex, and then
pointed out the relationship between conformal group and generation of mass
term. If only concerning mass-term problem in EoM, the structure group would
not be smaller than conformal group. Based on our analyses, the scaling
transformation allows a special chiral-like vertex $\gamma ^\mu (1\pm \gamma
_5)$ to be invariant. And normal vector vertex becomes running under such
scaling transformation. So with the scaling transformation only, one cannot
endow a massless fermion like ''neutrino'' with mass, since scaling
transformation happens to keep the chiral interaction form, and cannot make
fermions like ''neutrino'' deviate from the vertex form \cite{explaA}. So we
have to involve other generators of conformal group to transform the
interaction vertex, for example, the generator $K_\mu $. And coincidentally,
we find $K_\mu $ can also be responsible for transforming space-time from
space-like to time-like, which in physics should occur before transforming
fermions from massless to massive. With the above cognition, we get the mass
term by directly calculating the integral occurred in original paper, which
falls into one of the possibilities of our original surmise.

~~\\

According to the formula we get, we present a plausible understanding of the
generation of mass term. Since we have obtained the mass by directly
calculating the integral in original equation, we believe the mass term is
generated by the curvature (wrapped) of complex space. The wrapping of
complex space shows its effect by deforming electro-field and magnetic field
if we use the shortcut process $\gamma +\gamma \longrightarrow e^{+}e^{-}$
as paradigm. The progress implies that the photon in curved complex space is
split into particle pairs, and at the same time the electro-field and
magnetic-field of the photon are wrapped: the electro-field forms charge,
and magnetic field forms spin. This conclusion may help us make the mass
problem calculable, not just as legitimate parameters in DCSB and Higgs
mechanism. Of course it is necessary to confirm physically whether spin or
magnetic flux actually holds mass. And we should caution that here the
obtained mass may be just ''effective mass'', which roots obviously in
interaction. We conceive that this result might be helpful in understanding
spectra problem of hadrons, whose mass may be proportional to
(chromo-)magnetic flux. The further work along this line is in progress.

~~\\

We have achieved the conclusion that in our theory the minimal group should
be the entire conformal group. On mathematical side the conformal group has
been investigated thoroughly from different aspects \cite{Hor10}, and its
application to physics especially to quantum field once was also widely
considered. However the application is not so satisfactory because hitherto
no other perfect quantum system than photon field \cite{Bate1910,Cunn1910}
has been found so that the corresponding Lagrangian is conformally
invariant, unless the mass of the involved particles are null~\cite{Kast08,
Yu13, Jose88, Gross70, Dirac35,Lus75}. In our treatment we turn from
searching for invariant dynamics to how the conformal group makes physics
quantities running, so as to match the renormalization results that some of
the physics constants vary regularly with the energy scale, for instance
charge, mass, and Green functions. According to our analysis, the scaling
transformation can somehow cause variation of coupling constants and masses.
And in this paper, we have also discussed the function of generators $K_\mu $
and $P_\mu $ which conformally change the interaction vertex from a chiral
one to a non-chiral one. The action performed by generators $P_\mu $ or $%
K_\mu $ may be caused by the external non-hermite stimulation. Basically it
is with such stimulations that the value of $\vec E^2-\vec B^2$ changes from
$0$ to $\neq 0$. Then by certain decaying process, fermion mass is
generated. In summary, we get to the conclusion that the running of
interaction vertex corresponds to certain kinds of curving of complex space,
which is certainly governed by generators of conformal group. Conformal
group might live for curving, but not for invariance.

~~\\

\begin{acknowledgments}
I am grateful to Prof. W. Q. Wang and Prof. W. Han  for their
encouragements. The Project Sponsored by the Scientific Research
Foundation for the Returned Overseas Chinese Scholars, State
Education Ministry and Fundamental Research Funds for the Central
Universities, and Natural Scientific Foundation of China with
granted No.91227114.
\end{acknowledgments}

~~\\

\section{Appendix A. how $e^{\frac u2\gamma _5}$ exchange the fermion
spinors between helicity representation and spin representation}

We recognize here that the performance of $e^{\frac u2\gamma _5}$ can only
exchange the fermion spinors between helicity representation and spin
representation, or vice versa. Accordingly such transformation cannot yield
the generation of spinor, or the mass of fermions. In what follows we
present the example of how $e^{\frac u2\gamma _5}$ changes the
representation of spinors.

Despite of the algebric structure, while $m=0$ the spinor could be any
''vector-form'' homogenously because the term $\left(
\begin{array}{cccc}
a & b & c & d
\end{array}
\right) m\left(
\begin{array}{c}
a \\
b \\
c \\
d
\end{array}
\right) $ is equal to null. Nevertheless, when $m\neq 0$ , the term
spinor(vector) $\left(
\begin{array}{c}
a \\
b \\
c \\
d
\end{array}
\right) $ cannot be selected arbitrarily, since the condition $\left(
\begin{array}{cccc}
a & b & c & d
\end{array}
\right) m\left(
\begin{array}{c}
a \\
b \\
c \\
d
\end{array}
\right) =m$ should be satisfied. The most simple option is $\left(
\begin{array}{c}
a \\
b \\
c \\
d
\end{array}
\right) =\left(
\begin{array}{c}
1 \\
0 \\
0 \\
0
\end{array}
\right) $, or some analog like $\left(
\begin{array}{c}
0 \\
1 \\
0 \\
0
\end{array}
\right) $ can be used. Subsequently we will focus on the transformation from
the special spinor $\left(
\begin{array}{c}
1 \\
0 \\
0 \\
0
\end{array}
\right) $ to its helicity form in arbitrary reference coordinates.

Firstly, we refer to the Dirac equation for a rest particle
\begin{equation}
\gamma _0\psi =m\psi \text{ ,}  \label{A1}
\end{equation}
whose solution is $\left(
\begin{array}{c}
1 \\
0 \\
0 \\
0
\end{array}
\right) $ or $\left(
\begin{array}{c}
0 \\
1 \\
0 \\
0
\end{array}
\right) $ for normal particles, and $\left(
\begin{array}{c}
0 \\
0 \\
1 \\
0
\end{array}
\right) $ or $\left(
\begin{array}{c}
0 \\
0 \\
0 \\
1
\end{array}
\right) $\ for anti-particles. One also notes that $\left(
\begin{array}{cccc}
1 & 0 & 0 & 0
\end{array}
\right) \gamma _0\left(
\begin{array}{c}
1 \\
0 \\
0 \\
0
\end{array}
\right) =m$\ from Eq. (\ref{A1}). This knowledge legitimates the writing
from the rhs to lhs of Eq. (\ref{A1}). That means we can generate spinor via
mass term
\begin{equation}
m=\left(
\begin{array}{cccc}
1 & 0 & 0 & 0
\end{array}
\right) m\left(
\begin{array}{c}
1 \\
0 \\
0 \\
0
\end{array}
\right) =\left(
\begin{array}{cccc}
1 & 0 & 0 & 0
\end{array}
\right) \gamma _0\left(
\begin{array}{c}
1 \\
0 \\
0 \\
0
\end{array}
\right) \text{ .}  \label{A2}
\end{equation}
Assume that the motion of fermion is along z direction, one solution of
Dirac equation $(\not p-m)\psi =0$ in helicity representation is $\psi
=\left(
\begin{array}{c}
sth \\
0 \\
sth \\
0
\end{array}
\right) $. Further the transformation responsible for the variation from $%
\left(
\begin{array}{c}
1 \\
0 \\
0 \\
0
\end{array}
\right) $\ to $\left(
\begin{array}{c}
sth \\
0 \\
sth \\
0
\end{array}
\right) $ is $e^{\frac u2\gamma _5}$, explicitly
\begin{equation}
e^{\frac u2\gamma _5}\left(
\begin{array}{c}
1 \\
0 \\
0 \\
0
\end{array}
\right) =\left(
\begin{array}{c}
\cosh \frac u2 \\
0 \\
\sinh \frac u2 \\
0
\end{array}
\right) \text{ ,}  \label{A3}
\end{equation}
succeeding transformation $e^{\frac u2\gamma _5}$ could be employed if
necessary. Then the resultant spinor doesn't satisfy the Eq. (\ref{A1}) any
longer, which matches our common knowledge of quantum field theory.


\begin{thebibliography}{99}
\bibitem{explaA}  This is just a metaphor, in our next work it will be shown
that we actually want to resolve mass problem of fermion-like hadrons.
\end{thebibliography}
\end{document}